\documentclass[10pt]{article}

\usepackage{latexsym,amsmath,amssymb,epsfig}

\DeclareFontFamily{OT1}{rsfs10}{}
\DeclareFontShape{OT1}{rsfs10}{m}{n}{ <-> rsfs10 }{}
\DeclareMathAlphabet{\mathscript}{OT1}{rsfs10}{m}{n}

\numberwithin{equation}{section}

\newcommand{\be}{\begin{equation}}
\newcommand{\ee}{\end{equation}}
\newcommand{\nn}{\nonumber}
\newcommand{\bea}{\begin{eqnarray}}
\newcommand{\eea}{\end{eqnarray}}

\newcommand{\ns}{\normalsize}
\newcommand{\pt}{\partial}

\def\a{\alpha}
\def\b{\beta}
\def\g{\gamma}
\def\c{\chi}
\def\d{\delta}
\def\e{\epsilon}

\def\z{\psi}
\def\k{\kappa}
\def\l{\lambda}
\def\m{\mu}
\def\n{\nu}
\def\o{\omega}
\def\p{\pi}
\def\q{\theta}
\def\th{\theta}

\def\r{\rho}
\def\s{\sigma}
\def\t{\tau}

\def\x{\xi}
\def\z{\zeta}
\def\w{\wedge}

\def\G{\Gamma}

\def\L{\Lambda}
\def\O{\Omega}
\def\P{\Pi}
\def\T{\Theta}

\def\cA{{\cal A}}

\def\cF{{\cal F}}

\def\cG{{\cal G}}

\def\cC{{\cal C}}

\def\cX{{\cal X}}

\begin{document}


\begin{titlepage}

\vspace{-3cm}

\title{
   \hfill{\ns SUSX-TH/01-037\\}
   \hfill{\ns HUB-EP-01/34\\}
   \hfill{\ns hep-th/0109173}\\[3em]
   {\huge Five-Branes in Heterotic Brane-World Theories}\\[1em]}
   \setcounter{footnote}{0}
\author{
{\ns\large Matthias Br\"andle$^1$\footnote{email: brand@physik.hu-berlin.de}
  \setcounter{footnote}{3}
  and Andr\'e Lukas$^2$\footnote{email: a.lukas@sussex.ac.uk}} \\[0.8em]
   {\it\ns $^1$Institut f\"ur Physik, Humboldt Universit\"at}\\[-0.2em]
   {\ns Invalidenstra\ss{}e 110, 10115 Berlin, Germany}\\[0.4em]
   {\it\ns $^2$Centre for Theoretical Physics, University of Sussex}
   \\[-0.2em]
   {\ns Falmer, Brighton BN1 9QJ, UK} \\[0.2em] }   
\date{}

\maketitle

\begin{abstract}
The effective action for five-dimensional heterotic M-theory in
the presence of five-branes is systematically derived from
Ho\v rava-Witten theory coupled to an M5-brane world-volume theory.
This leads to a five-dimensional $N=1$ gauged supergravity
theory on $S^1/\mathbb{Z}_2$ coupled to four-dimensional
$N=1$ theories residing on the two orbifold fixed planes and
an additional bulk three-brane. We analyse the properties of this
action, particularly the four-dimensional effective theory associated
with the domain-wall vacuum state. The moduli K\"ahler potential
and the gauge-kinetic functions are determined along with the
explicit relations between four-dimensional superfields and
five-dimensional component fields.
\end{abstract}

\thispagestyle{empty}

\end{titlepage}


\section{Introduction}

A large class of attractive five-dimensional brane-world models can be
constructed by reducing Ho\v rava-Witten
theory~\cite{Horava:1996qa,Horava:1996ma,Witten:1996mz} on
Calabi-Yau three-folds. This procedure has been first carried out in
Ref.~\cite{Lukas:1999yy,Lukas:1999tt,Ellis:1999dh} and it leads to gauged
five-dimensional $N=1$ supergravity on the orbifold $S^1/\mathbb{Z}_2$
coupled to $N=1$ gauge and gauge matter multiplets located on the two
four-dimensional orbifold fixed planes. It has been
shown~\cite{Andreas:1999ei}--\cite{Donagi:2000zs}
that a phenomenologically interesting particle spectrum on the orbifold
planes can be obtained by appropriate compactifications.

Early on it has been realized~\cite{Witten:1996mz} that M5-branes being
transverse to the orbifold direction, wrapping a holomorphic curve in
the Calabi-Yau space and stretching across the four uncompactified
dimensions can be incorporated into this picture. The explicit form of
the corresponding 11-dimensional vacuum solutions has been given in
Ref.~\cite{Lukas:1999hk}. In the five-dimensional brane-world theory
such an M5-brane appears as a three-brane located in the bulk away
from the orbifold fixed planes.  This provides an interesting
generalisation of five-dimensional heterotic brane-world models which
has recently attracted some
attention~\cite{Khoury:2001wf}--\cite{Arnowitt:2001ti}, particularly in
the context of cosmology.  Some features of such generalised
brane-world models have already been analysed in
Ref.~\cite{Lukas:1999hk} and used in subsequent applications. The
purpose of this paper is to present a systematic derivation of the
five-dimensional effective action for these models in its simplest
form and discuss its properties.

Starting point for this derivation is Ho\v rava-Witten theory in 11
dimensions coupled to an M5-brane world-volume
theory~\cite{Bandos:1997ui}. The action for this system is obtained by
combining results presented in
Refs.~\cite{Horava:1996ma,deAlwis:1997gq,Bandos:1998gd}. We show
explicitly that the warped vacua based on Calabi-Yau three folds
presented in Ref.~\cite{Lukas:1999hk} can be promoted to solutions of
this coupled theory. Given previous results, this amounts to showing
that the five-brane sources in the Einstein equation are properly
matched and that the five-brane world-volume equations of motion are
satisfied.  We obtain the five-dimensional brane-world theory by
performing a reduction on a Calabi-Yau three-fold focusing on the
universal sector of the Calabi-Yau zero modes. The result is
consistent with previous expectations~\cite{Lukas:1999hk} and
represents a five-dimensional $N=1$ gauged supergravity on
$S^1/\mathbb{Z}_2$ coupled to the $N=1$ theories on the orbifold
planes and an additional $N=1$ theory located on the three-brane. We
verify that the BPS domain-wall of Ref.~\cite{Lukas:1999yy} can be
generalised to include the effect of this additional three-brane. The
action, as obtained by reduction from 11 dimensions, is
expressed in terms of the ``step-function'' $\a$ which represents the
mass parameter of the gauged supergravity. For this form of the
action, the three-brane is coupled magnetically. We also present a
dual action where $\a$ is replaced by a four-form with five-form field
strength to which the three-brane couples electrically.  Finally, we
analyse the four-dimensional $N=1$ effective theory associated to the
domain-wall vacuum state. We determine the moduli K\"ahler potential
for the dilaton, the $T$-modulus and the five-brane position
modulus. Our result agrees with Ref.~\cite{Derendinger:2001gy} where
somewhat different methods have been employed. In addition, we obtain
the explicit relations between the four-dimensional superfields and
the five-dimensional component fields which are vital whenever
four-dimensional results have to be interpreted in terms of the
five-dimensional brane-world theory. We also compute the gauge-kinetic
functions for the orbifold gauge fields and find threshold corrections
in agreement with
Ref.~\cite{Lukas:1999hk,Derendinger:2001gy}. Finally, we show that the
gauge-kinetic function for the three-brane gauge fields does not
receive any threshold corrections at leading order and is simply given
by the period matrix of the complex curve wrapped by the five-brane.


\section{M5-brane coupled to D=11 Supergravity on an Orbifold}

\subsection{The 11-dimensional action}

In this section, we would like to set the stage by describing our
starting point, the effective $D=11$ action for
M-theory on the orbifold $S^1/\mathbb{Z}_2$ coupled to a
five-brane. In order to derive this action, we can draw information
from two main sources, namely the Ho\v{r}ava--Witten (HW)
action~\cite{Horava:1996qa,Horava:1996ma} for M-theory on
$S^1/\mathbb{Z}_2$ and the action for $D=11$ supergravity coupled to a
five-brane due to Bandos, Berkovits and
Sorokin~\cite{Bandos:1998gd,deAlwis:1997gq}. Combining these two
results it not completely straightforward and requires a careful
analysis of the various symmetries involved. A detailed account of
this will be given in a forthcoming work~\cite{Inprep}. Here we will
merely present the final result for the bosonic part of this action
which reads~\footnote{For the bulk fields we adopt the normalisation
of Ref.~\cite{Bandos:1998gd}. The normalisation chosen by
Ho\v rava and Witten~\cite{Horava:1996ma} is obtained by the rescaling
$g_{\rm HW}=2^{-2/9}g$, $C_{\rm HW}=\frac{2^{1/6}}{6}C$ and
$G_{\rm HW} = 2^{1/6}G$.}
\begin{eqnarray}\label{action}
  S &=& -\frac{1}{2\k^{2}}\int_{M}\left\{d^{11}x\,
        \sqrt{-g}\,\left(\frac{1}{2}R+\frac{1}{4!}G_{IJKL}G^{IJKL}\right)
                +\frac{2}{3}C\w \cG\w \cG\right\}\nn\\
     &&-\frac{1}{4\lambda^{2}} \sum_{k=1}^{2} \int_{M_{10}^{k}}
       d^{10}x \sqrt{-g_{10}}\left\{{\rm tr}\,F_{k}^{2}-
       \frac{1}{2}{\rm tr}\,R^{2}\right\}\nn\\   
     &&-\frac{1}{2}T_{5}\int_{M_{6}\cup\tilde{M}_6}d^{6}\s
       \sqrt{-\g}\left[1+v_{l}(*H)^{lmn}(*H-H)_{mnp}v^{p}\right]
       +2\,dB\w \hat{C}\\
 && +T_{5}\int_{M}C\w dC\w\left[ \T (M_{6})+\T (\tilde{M}_{6})\right]\nn\; .
\end{eqnarray}
The structure of 11-dimensional space-time in this action is
$M=M_{10}\times S^{1}/\mathbb{Z}_{2}$ where $M_{10}$ is
ten-dimensional space-time and we work in the upstairs picture. As
usual, we define the orbifold coordinate $y=x^{11}$ to be in the range
$y\in[-\p\r,\p\r]$ and let the $\mathbb{Z}_2$ orbifold symmetry act
as $y\longrightarrow -y$.  This leads to the two fixed ten-dimensional
hyperplanes $M_{10}^{1}$ and $M_{10}^{2}$ located at $y=y_1=0$ and
$y=y_2=\p\r$, respectively. Further, we have a single
five-brane~\footnote{The generalisation to include an arbitrary number
of five-branes is straightforward and will be covered in
Ref.~\cite{Inprep}.} with world-volume $M_6$ plus its $\mathbb{Z}_2$
mirror with world-volume $\tilde{M}_6$ which originates from $M_6$ by
applying the orbifold map $y\longrightarrow -y$. This latter mirror
five-brane is required by consistency in order to keep the theory
$\mathbb{Z}_2$ symmetric. Further, to avoid the appearance of
additional states~\cite{Ganor:1996mu}, we demand that the five-brane
world-volume does not intersect either of the two orbifold fixed
planes. We use indices $I,J,K,\ldots = 0,\ldots ,10,11$ for
11-dimensional space-time with coordinates $x^I$ and indices
$m,n,p,\ldots = 0,\ldots ,5$ for the five-brane world-volume with
coordinates $\s^m$.

Let us now discuss the various sectors of the above action in some
detail. The bulk fields consist of the fields of 11-dimensional
supergravity, that is the $\mathbb{Z}_2$--even~\footnote{We call a
tensor field $\mathbb{Z}_2$--even if its components orthogonal to the
orbifold are even, otherwise we call it $\mathbb{Z}_2$--odd.}
11-dimensional metric $g_{IJ}$, the $\mathbb{Z}_2$--odd three-index
antisymmetric tensor field $C_{IJK}$ and the gravitino $\Psi_I$,
subject to the usual $\mathbb{Z}_2$
truncation~\cite{Horava:1996ma}. The standard relation $G=dC$ between
$C$ and its field strength $G$ will be modified due to the presence of
source terms and this will be explicitly presented shortly. Anomaly
cancellation requires the two orbifold fixed planes $M_{10}^{k}$ to
each carry a 10-dimensional $N=1$ $E_8$ gauge
multiplet~\cite{Horava:1996ma}, that is an $E_8$ gauge field $A_k$
with field strength $F_k$ and gauginos $\c_k$, where $k=1,2$. The
Yang-Mills coupling $\l$ is fixed in terms of the 11-dimensional
Newton constant $\k$
by~\cite{Horava:1996ma,Conrad:1998ww,Harmark:1998bs}
\begin{equation}
  \l^{2}=4\p(4\p\k^{2})^{\frac{2}{3}}\; .
\end{equation}
The five-brane world-volume fields consist of the embedding coordinates
$X^I=X^I(\s^m)$ together with the fermions $\th$ and the two-index
antisymmetric tensor field $B_{mn}$. The five-brane part of the above
action is written in the form due to Pasti, Sorokin and Tonin
(PST)~\cite{Pasti:1997gx} which requires the introduction of an
auxiliary scalar field $a$ and an associated unit vector field $v_m$
defined by
\begin{equation}
  v_{m}=\frac{\pt_{m}a}{\sqrt{\g^{np}\pt_{n}a\pt_{p}a}}\; .
\end{equation}
The presence of this field enhances the symmetries of the action
such that $a$ is truly auxiliary and that fixing one
of the symmetries turns the equation of motion for $B$ into the
self-duality condition $*H=H$. For this to actually work the Wess-Zumino term 
$dB\w\hat{C}$ must be present. For simplicity, we have chosen to present
a linearised form of the PST action as is appropriate for our subsequent
discussion. Here we will not consider any further details of
the PST-formulation and how precisely it relates to the derivation of our
action~\eqref{action}, but instead refer to Ref.~\cite{Inprep}
for a detailed discussion. As usual, the metric $\g_{mn}$ is the pull-back
\begin{equation}
 \g_{mn} =\pt_mX^I\pt_nX^Jg_{IJ}\; 
\end{equation}
of the space-time metric $g_{IJ}$. Further, the field strength $H$ of
$B$ is defined by
\begin{equation}
 H = dB-\hat{C}
\end{equation}
where $\hat{C}$ denotes the pull-back of the bulk field $C$, that is,
\begin{equation}
 \hat{C}_{mnp}=\pt_{m}X^{I}\pt_{n}X^{J}\pt_{p}X^{K}C_{IJK}\; .
\end{equation}
The five-brane tension $T_5$ can be expressed in terms of the
11-dimensional Newton constant as
\begin{equation}
T_{5}=\left(\frac{\p}{2\k^{4}}\right)^{\frac{1}{3}}\; .
\end{equation}
Having introduced all fields we should now specify the source terms
in the definition of the bulk antisymmetric tensor field strength.
To this end, we introduce
\bea
 G &=& dC-\o_{\rm YM}-\o_{\rm M5} \label{G}\\
 \cG &=& dC-\o_{\rm YM}\; .
\eea
The field strength $\cG$ is defined as in pure HW theory without five-branes,
that is, it only contains the ``Yang-Mills'' sources $\o_{\rm YM}$
which originate from the orbifold fixed planes and are given
by~\footnote{By $\d (y)$ we denote a $\d$--function one-form defined by
$\hat{\d}(y)dy$, where $\hat{\d}(y)$ is the ordinary $\d$--function.}
\begin{equation}
 \o_{\rm YM} = 2k\left[\o_1\wedge\d (y)+\o_2\wedge\d (y-\p\r )\right]
\end{equation}
with the ``Chern-Simons'' forms $\o_k$ satisfying
\begin{equation}
 J_k\equiv d\o_k = \frac{1}{16\p^{2}}\left[{\rm tr}\,F_{k}\w F_{k}-
                   \frac{1}{2}{\rm tr}\, R\w R\right]_{y=y_{k}}\; ,
\end{equation}
where $k=1,2$. The field strength $G$, on the other hand, contains both
orbifold and five-brane sources where the latter are defined by
\begin{equation}
 \o_{\rm M5} = k\left[\T (M_6)+\T (\tilde{M}_6 )\right]\; .
\end{equation}
Here $\T (M_6)$ is the $\q$--function associated with the five-brane
world-volume $M_6$. In analogy with the ordinary one-dimensional
$\q$--function it satisfies the relation
\begin{equation}
 d\T (M_6)= \d (M_6)\; ,
\end{equation}
where $\d (M_6)$ is the $\d$--function supported on $M_6$ (Analogous
expression hold for $\tilde{M}_6$.). For later calculations it will be
useful to explicitly express these functions in terms of the embedding
coordinates $X^I$ by writing
\begin{eqnarray}
  \T(M_{6})&=&\frac{1}{4!7!\sqrt{-g}}\,dx^{I_{1}}\w\ldots\w
               dx^{I_{4}}\e_{I_{1}\ldots I_{11}}\int_{M_{7}} dX^{I_{5}}
               \w\ldots\w dX^{I_{11}}\,\hat{\d}^{11}(x-X(\s)) \label{Theta}\\
  \d(M_{6})&=&\frac{-1}{5!6!\sqrt{-g}}\,dx^{I_{1}}\w\ldots\w
              dx^{I_{5}}\e_{I_{1}\ldots I_{11}}\int_{M_{6}=\pt M_{7}}
              dX^{I_{6}}\w\ldots\w dX^{I_{11}}\,\hat{\d}^{11}(x-X(\s))\; .
              \label{delta}
\end{eqnarray}
We see that the definition of $\T (M_6)$ and, hence, our
action~\eqref{action} involves a seven-manifold $M_7$ which is bounded
by the five-brane world-volume $M_6$, that is, $\pt M_7=M_6$. This
seven-manifold is the analogue of a Dirac-string for a monopole in
Maxwell theory and is also referred to as
Dirac-brane~\cite{Lechner:2001pn,Lechner:2001sj}. There may be a
problem in that the action depends on the particular choice of the
Dirac-brane. A prescription to resolve this ambiguity has been
proposed in ref.~\cite{deAlwis:1997gq}. Since our subsequent
considerations do not depend on how precisely the Dirac-brane is
defined we will not consider this point in any further detail. The
constant $k$ in the above definitions for the field strengths is again
fixed in terms of the 11-dimensional Newton constant and is given by
\begin{equation}
 k=\left(\frac{\p}{2}\right)^{\frac{1}{3}}\k^{2/3}
  =\k^{2}T_{5}
  =8\p^{2}\,\k^{2}/{\l^{2}}\; .
\end{equation}
From the definition~\eqref{G} we can now write the Bianchi identity
\begin{equation}
  dG = -2k\left[J_{1}\wedge\d(y)+J_{2}\wedge\d(y-\p\r)+
        \frac{1}{2}\left(\d(M_{6})
        +\d(\tilde{M}_{6})\right)\right],
 \label{BI}
\end{equation}
for $G$ which will be important later on. The relative factor $1/2$ between
the orbifold and five-brane sources accounts for the fact that the
five-brane and its mirror really represent the same physical object and
should, therefore, not be counted independently. 

\subsection{Symmetries}

Let us discuss the symmetries of the action~\eqref{action} some of which
will become relevant later on.

In the following we would like to check the BPS property of
certain solutions, hence we will need the (bosonic part) of the supersymmetry
transformations which we explicitly present for completeness. For the
gravitino $\Psi_I$, the $E_8$ gauginos $\c_k$ on the two orbifold
fixed planes and the five-brane world-volume fermions $\th$ they are,
respectively, given by
\bea
 \d\Psi_{I} &=& D_{I}\eta+\frac{1}{3!}\left[\frac{1}{4!}
                \G_{IJ_{1}J_{2}J_{3}J_{4}}-\frac{2}{3!}g_{IJ_{1}}
                \G_{J_{2}J_{3}J_{4}}\right]
                G^{J_{1}J_{2}J_{3}J_{4}}\eta \label{Psi}\\
 \d\c_k &=& -\frac{1}{4}\G^{\bar{I}\bar{J}}F_{k\bar{I}\bar{J}}\eta
             \label{chi}\\
 \d\th  &=& \eta+P_{+}\k \label{theta}
\eea
where the projection operators $P_\pm$ satisfying $P_++P_-=1$ are defined
by
\begin{equation}
  P_{\pm}=\frac{1}{2}\left(1\pm\e^{m_{1}\ldots
               m_{6}}\pt_{m_{1}}X^{I_{1}}\ldots\pt_{m_{6}}X^{I_{6}}
             \G_{I_{1}\ldots I_{6}}\right).\label{P}
\end{equation}
For simplicity, we have stated these projection operators for the
later relevant case $H=0$. The general expressions can be found in
Ref.~\cite{Claus:1998cq}. In the above equations the spinor $\eta$
parametrises supersymmetry transformations. The five-brane world-volume
theory is also invariant under an additional fermionic symmetry, namely
local $\k$--symmetry. It is parametrised by the spinor $\k$ and
appears via the second term in Eq.~\eqref{theta}. Further, $D_{I}$ is
the covariant derivative and $\G_{I_1\ldots I_p}$ denotes the
antisymmetrised products of $p$ gamma-matrices $\G_I$ which satisfy
the usual Clifford algebra $\{\G_{I},\G_{J}\}=2g_{IJ}$.

\vspace{0.4cm}
      
Besides supersymmetry the action is also, up to total derivatives, invariant
under the following gauge variations
\begin{equation}\label{gausym1}
\d C=d\L^{(2)},\qquad\d B=d\L^{(1)}+\hat{\L}^{(2)},
\end{equation}
where $\L^{(1)}$ is an arbitrary one-form. The two-form $\L^{(2)}$ has
to be $\mathbb{Z}_2$--odd in order to ensure that the $\mathbb{Z}_{2}$
properties of $C$ are preserved under the above transformation.

\vspace{0.4cm}

There are two more symmetries on the world-volume of the M5-brane, namely
the ``PST-symmetries'' given by
\begin{equation}
  \d B_{mn}=(da\w\phi^{(1)})_{mn}-
            \frac{\varphi}{\sqrt{(\pt a)^{2}}}\,v^{l}(*H-H)_{lmn},
            \qquad\quad
  \d a=\varphi \label{PST}
\end{equation}
where $\phi^{(1)}$ and $\varphi$ are an arbitrary one-form and a scalar, 
respectively. As previously mentioned, these symmetries ensure that
the self-duality of $H$ follows from the equations of motion and that
$a$ is an auxiliary field.


\section{Calabi-Yau background in $D=11$}\label{sectsol}

\subsection{The solution}

Background solutions of heterotic M-theory based on Calabi-Yau three-folds
which respect four-dimensional Poincar\'e invariance and $N=1$
supersymmetry were first presented in Ref.~\cite{Witten:1996mz}.
This paper also demonstrated how to include five-branes in those backgrounds
while preserving the four-dimensional symmetries. The explicit form of these
solutions was subsequently given in Ref.~\cite{Lukas:1999hk}. All these
result were based on the original action derived by Ho\v rava and
Witten~\cite{Horava:1996ma} which does not explicitly include any five-brane
world-volume theories. The effect of five-branes on the supergravity
background was incorporated by modifying the Bianchi-identity of $G$
to include the five-brane sources as in Eq.~\eqref{BI}. The main
purpose of this section is to proove that the solutions obtained in this
way can indeed be extended to solutions of the full action~\eqref{action}
which does include the five-brane world-volume theory. Practically, this
amounts to showing that these solutions correctly match the five-brane source
terms in the Einstein equations and that the five-brane world-volume equations
of motion are satisfied.

\vspace{0.4cm}

Following Ref.~\cite{Lukas:1999hk}, let us start reviewing the
solutions which are constructed as an expansion in powers of
$\k^{2/3}$.  At lowest order, we consider the space-time structure
$M=S^{1}/\mathbb{Z}_{2}\times X\times M_{4}$, where $X$ is a
Calabi-Yau three-fold and $M_{4}$ four-dimensional Minkowski
space. Coordinates in $M_4$ are labelled by indices $\m ,\n ,\r
,\ldots = 0,\ldots ,3$. The Ricci-flat metric on the Calabi-Yau space
is denoted  by $\O_{AB}$ with six-dimensional indices $A,B,C,\ldots
=5,\ldots ,10$. The K\"ahler-form $\o$ is defined by
$\o_{a\bar{b}}=i\O_{a\bar{b}}$ where $a,b,c,\ldots$ and
$\bar{a},\bar{b},\bar{c},\ldots$ are holomorphic and anti-holomorphic
indices on the Calabi-Yau space, respectively. For simplicity, we will
restrict our considerations to the universal sector of the Calabi-Yau
space, that is, strictly our results apply to Calabi-Yau spaces with
$h^{1,1}=1$. The four-form field strength $G$ vanishes at lowest
order. This configuration constitutes a solution to the Killing spinor
equation $\d\Psi_I=0$ and the Bianchi identity since the source terms
in Eq.~\eqref{BI} are proportional to $\k^{2/3}$ and, hence, do not
contribute at lowest order. At the next order, however, these source
terms have to be taken into account and, as a consequence, the field
strength $G$ becomes non-vanishing. This induces corrections to the
metric which can be computed requiring that $N=1$ supersymmetry is
preserved and, hence, that the gravitino variation~\eqref{Psi}
vanishes. The size of these corrections is measures by the
strong-coupling expansion parameter $\e_S$ defined by
\begin{equation}
 \e_S \equiv \p\left(\frac{\k}{4\p}\right)^{2/3}\frac{2\p\r}{v^{2/3}}
      = \p\r T_5\k^2\label{eS}
\end{equation}
where $v=\int_X\sqrt{\O}$ is the Calabi-Yau volume.

We should now specify the full solutions (to order $\e_S$) and we
start with the gauge fields on the orbifold planes. In general, we
have non-trivial holomorphic vector bundles on the Calabi-Yau
space. These bundles correspond to gauge field backgrounds $\bar{A}_k$
in the Calabi-Yau directions which preserve supersymmetry and are,
hence, constrained by a vanishing gaugino variation~\eqref{chi}. This
implies that their associated field strengths $\bar{F}_k$ are $(1,1)$ forms
on the Calabi-Yau space. Then, the orbifold sources
$J_k$ in the Bianchi-identity are $(2,2)$ forms given by
\begin{equation}
 J_k\equiv d\o_k = \frac{1}{16\p^{2}}\left[{\rm tr}\,\bar{F}_{k}\w
                   \bar{F}_{k}-\frac{1}{2}{\rm tr}\, R^{(\O )}\w R^{(\O )}
                   \right]_{y=y_{k}}\; ,
\end{equation}
where $R^{(\O )}$ is the Calabi-Yau curvature tensor associated with
the metric $\O$.

Next, we should consider the five-brane world-volume fields. Guided by
the structure of our action~\eqref{action}, we focus on a single
five-brane (and its $\mathbb{Z}_{2}$ mirror) which is taken to be
static and parallel to the orbifold fixed planes. Furthermore, two
spatial dimensions of the world-volume $M_{6}$ wrap around a
holomorphic two-cycle $\cC_{2}$ of the internal Calabi-Yau space $X$
and the remaining four dimensions stretch across the external
Minkowski space-time $M_4$. Accordingly, we split the five-brane
coordinates $\s^m$ into external and internal coordinates, that is,
$\s^m = (\s^\m ,\s^i)$ where $\m ,\n ,\r ,\ldots = 0,\ldots ,3$ and
$i,j,\ldots = 4,5$. Further, we define holomorphic and
anti-holomorphic coordinates $\s = \s^4 +i\s^5$ and $\bar{\s} = \s^4
-i\s^5$. With these definitions, the five-brane embedding is specified
by
\begin{equation}
 X^\m = \s^\m\, ,\qquad X^a = X^a(\s )\, ,\qquad X^{11}=\pm Y
 \label{emb}
\end{equation}
where $Y\in [0,\p\r ]$ is a constant, $X^a(\s )$ parametrises the
holomorphic curve $\cC_2$ and the two signs in the last equation account
for the five-brane $M_6$ and its mirror $\tilde{M}_6$. The world-volume
two-form $B$ is taken to vanish in the background. It can be explicitly
shown~\cite{Lukas:1999hk} that this configuration preserves supersymmetry
on the five-brane by choosing $\k = -\eta$ in the variation~\eqref{theta}
and verifying that $P_-\eta = 0$. The five-brane source in the
Bianchi identity then takes the specific form
\begin{eqnarray}\label{M5solcur}
  J_{M5}&\equiv&d\o_{\rm M5}=kJ_5\wedge\left[\d(y-Y)
         +\d(y+Y)\right]\\
  J_{5}&=&\d(\cC_{2})=\frac{1}{2\cdot 4!\sqrt{\O}}dx^{A_{1}}\w\ldots\w
         dx^{A_{4}}\e_{A_{1}\ldots A_{4}BC}\int_{\cC_{2}}dX^{B}\w
         dX^{C}\hat{\d}^{6}(x-X(\s))\; .
\end{eqnarray}
The embedding~\eqref{emb} implies that $J_5$ is a $(2,2)$ form on the
Calabi-Yau space as well.

Now we have explicitly presented all source terms in the Bianchi identity
and we can use Eqs.~\eqref{BI} and \eqref{Psi} to determine $G$ and the
corrected metric. The result is 
\begin{eqnarray}
   ds_{11}^{2} &=& (1-h)\eta_{\m\n}dx^{\m}dx^{\n}+(1+h)\O_{AB}dx^{A}dx^{B}
                   +(1+2h)dy^{2} \label{gsol} \\
   G &=& \frac{1}{2}\pt_{y}h(y)*\o\label{Gsol}
\end{eqnarray}
where
\begin{equation}\label{bfun}
  h(y)=-\frac{2}{3}\left\{ \begin{array}{lll}
      \a_{1} |y|+c &\mbox{for}& 0\leq |y|\leq Y \\
      (\a_{1}+\a_{5})|y|-\a_{5}\, Y+c &\mbox{for}&
      Y\leq |y|\leq \p\r
      \end{array}\right.
\end{equation}
where $c$ is a constant and the charges $\a_k$ are defined by
\begin{equation}
 \a_k = \frac{\e_S}{\p\r}\b_k\, , \qquad \b_k = \int_X\o\wedge J_k
 \label{alphak}
\end{equation}
for $k=1,2,5$. From the Bianchi identity~\eqref{BI} these charges must
satisfy the cohomology condition
\begin{equation}
\a_1+\a_2+\a_5 = 0\; .
\end{equation}

\vspace{0.4cm}

We would now like to demonstrate that the above configuration which
was mainly obtained by requiring unbroken $N=1$, $D=4$ supersymmetry
is indeed a solution of the equations of motion derived from the
action~\eqref{action}. Given previous results~\cite{Lukas:1998fg}, what
remains to be shown is that the five-brane sources in the Einstein
equation are properly matched and the five-brane world-volume equations
of motion are satisfied. To verify the former we should consider the
singular terms in the Einstein tensor which turn out to be
\begin{equation}
 (G_{\m\n})_{\rm singular}=-\frac{3}{2}\pt_y^2h\,\eta_{\m\n}\, ,\qquad
 (G_{AB})_{\rm singular}=-\frac{1}{2}\pt_y^2h\,\O_{AB} \label{GG}\, .
\end{equation}
These terms have to be compared with the five-brane stress energy
tensor which, in general, is given by
\begin{equation}
 T_{5IJ} = T_5\k^2\frac{1}{\sqrt{-g}}\int_{M_6\cup\tilde{M}_6}
           d^6\s\hat{\d}^{11}(x-X(\s ))
           \sqrt{-\g}\g^{mn}\pt_mX_I\pt_nX_J\, .
\end{equation}
Evaluating this expression for the embedding~\eqref{emb} leads to
\begin{equation}
 T_{5\m\n}=T_5\k^2\b_5\eta_{\m\n}\left(\d (y-Y)+\d (y+Y)\right)\, ,\qquad
 T_{5AB}=\frac{1}{3}T_5\k^2\b_5\O_{AB}\left(\d (y-Y)+\d (y+Y)
         \right)\label{Tsol}
\end{equation}
with the other components vanishing. In view of Eqs.~\eqref{bfun} and
\eqref{eS} this exactly matches the appropriate delta-function terms
in eq.~\eqref{GG}. We have therefore verified that the five-brane
sources in the Einstein equation are properly matched by the
solutions.

The only relevant equation of motion on the five-brane world-volume
is the one for the embedding coordinates $X^I$. For the case of
vanishing $B$ it reads
\begin{equation}
 \Box X^I+\G_{JK}^I\g^{mn}\pt_mX^J\pt_nX^K+\frac{2}{6!}\e^{m_1\ldots m_6}
 \pt_{m_1}X^{I_1}\ldots \pt_{m_6}X^{I_6}{(*G)^I}_{I_1\ldots I_6}=0\; .
\end{equation}
The $\m$ and $A$ components of this equation turn out to be trivially
satisfied for our solution and it remains to check the $11$ component.
Using the expressions
\begin{equation}
 \G_{\m\n}^{11}=\frac{1}{2}\pt_yh\eta_{\m\n}\, ,\qquad
 \G_{AB}^{11}=-\frac{1}{2}\pt_yh\O_{AB}
\end{equation}
for the connection along with the embedding~\eqref{emb} and the
background~\eqref{Gsol} for $G$ this can indeed easily be done.

In summary, we have, therefore, explicitly verified that the
above background configurations are indeed solutions of the
action~\eqref{action}.

\subsection{Including moduli}

In view of the reduction to five dimensions to be carried out shortly
we will now identify the (bosonic) moduli fields of the above
background solutions. We will use indices $\a ,\b ,\g =0,\ldots ,3,11$
to label five-dimensional coordinates.

Let us start with the bulk fields. As mentioned earlier, we focus on
the universal Calabi-Yau sector for simplicity, that is, we are
considering Calabi-Yau spaces with $h^{1,1}=1$. The general case
will be examined in Ref.~\cite{Inprep}. By absorbing the
corrections into the five-dimensional moduli as explained in
Ref.~\cite{Lukas:1999yy} the metric can be written as
\begin{equation}\label{metans}
  ds_{11}^{2}=V^{-2/3}g_{\a\b}dx^{\a}dx^{\b}+V^{1/3}\O_{AB}dx^{A}dx^{B}\; .
\end{equation}
where the Calabi-Yau volume modulus $V$ and the five-dimensional metric
$g_{\a\b}$ are functions of the five-dimensional coordinates $x^\a$.
From the bulk antisymmetric tensor field $C$ we have, in five
dimensions, a three form $C_{\a\b\g}$ with field strength $G_{\a\b\g\d}$,
a vector field $\cA_{\a}$ with field strength $\cF_{\a\b}$ and a complex
scalar field $\x$ with field strength $\cX_{\a}$. These fields are
defined by
\begin{equation}
\label{Cmoduli}
\begin{array}{lll}
  C_{\a\b\g}&\qquad&G_{\a\b\g\d}=4\pt_{[\a}C_{\b\g\d]}\nn \\
  C_{\a AB}=\cA_{\a}\,\o_{AB}&\qquad&G_{\a\b AB}=\cF_{\a\b}\,\o_{AB}\\
  C_{ABC}=\xi\,\o_{ABC}&\qquad&G_{\a ABC}=\cX_{\a}\,\o_{ABC} 
\end{array}
\end{equation}
where $\o_{ABC}$ is the harmonic $(3,0)$-form on the Calabi-Yau space. 

We now turn to the boundary theories. We have already mentioned that,
on both boundaries, we have internal gauge bundles on the Calabi-Yau
space. The external parts of the gauge fields, denoted by $A_{k\m}$
with field strengths $F_{k\m\n}$, lead to gauge fields on the now
four-dimensional orbifold fixed planes $M_{4}^{k}$, where $k=1,2$.
Their gauge groups are given by the commutants of the
internal structure group within $E_8$. For simplicity, we will not
consider any gauge matter fields on $M_4^k$, although they will be
included in \cite{Inprep}.

Next we should discuss the zero modes on the five-brane world-volume.
The five-brane is allowed to fluctuate in five external dimensions, while
internally it can move within the Calabi-Yau space. This leads to the
following set of embedding coordinates
\begin{equation}
  X^{\mu}=X^{\mu}(\s^{\nu})\, ,\qquad X^{11}=Y(\s^{\nu})\, ,\qquad
  X^{a}=X^{a}(\s ,M)
\end{equation}
where $M$ is a set of moduli which parametrises the moduli space of
holomorphic curves with a given homology class $[\cC_2]$ for the
Calabi-Yau space under
consideration~\cite{Donagi:1999xe,Donagi:1999gc,Donagi:1999jp}. In our
low-energy effective action, we will not explicitly take these moduli
into account.  The three-brane surface in five-dimensional space specified
by the above embedding coordinates $X^\a$ is denoted by $M_4^5$. Using
the above embedding and the bulk metric (\ref{metans}) we find the
following non-vanishing components of the induced world-volume metric
\begin{eqnarray}
 \g_{\mu\nu} &=& \pt_{\mu}X^{\a}\pt_{\nu}X^{\b}g_{\a\b}\\
  \g_{jk} &=& \pt_{j}X^{A}\pt_{k}X^{B}\O_{AB}=
    2\pt_{\s}X^{a}\pt_{\bar{\s}}X^{\bar{b}}\O_{a\bar{b}}\,\d_{jk}\; .
\end{eqnarray}
There are also a number of moduli arising from the two-form $B$
which can be determined from the cohomology of the two-cycle
$\cC_2$. We introduce a basis $\l_U$ of $H^1(\cC_2)$ where
$U,V,W,\ldots =1,\ldots ,2g$ and $g$ is the genus of $\cC_2$ while the
pull-back $\hat{\o}$ of the Calabi-Yau K\"{a}hler form to the two-cycle
$\cC_2$ provides a basis for $H^2(\cC_2)$. Then we find the two-form
$B_{\m\n}$ with field strength $H_{\m\n\r}$, $2g$
Abelian~\footnote{This can be enhanced to non-abelian symmetries if
five-branes are ``stacked'', as discussed in
Ref.~\cite{Lukas:1999hk}. We do not attempt to incorporate this effect
explicitly.} vector fields $D^U_\m$ with field strengths $E^U_{\m\n}$
and a scalar $s$ with field strength $j_\m$ as the low-energy fields
on the three-brane $M_4^5$. These fields are defined by
\begin{equation}
\begin{array}{lll}
  B_{\mu\nu}&\qquad&H_{\mu\nu\r}=(dB-\hat{C})_{\mu\nu\r}\\
  B_{\mu j}=D^{U}_{\mu}\,\l_{U\,j}&\qquad &
  H_{\mu\nu j}=E^{U}_{\mu\nu}\l_{U\,j}=(dD^{U})_{\mu\nu}\l_{U\,j}\\
  B_{jk}=s\,\hat{\o}_{jk}&\qquad &
  H_{\mu jk}=j_{\mu}\hat{\o}_{jk}=(ds-\hat{\cA})_{\mu}\hat{\o}_{jk}
\end{array}
\end{equation}
Due to the self-duality condition $*H=H$ these four-dimensional fields
are not all independent. In order to work out the relations between them
we split the $2g$ vector fields into two sets, that is,
we write $(E^U)=(E^u,\tilde{E}_u)$ where $u,v,w,\ldots = 1,\ldots ,g$.
Then, we find that the self-duality condition reduces to
\begin{eqnarray}
  j&=&V\,*H \label{dualj}\\
  \tilde{E}_{v}&=&[{\rm Im}(\Pi)]_{vw}*E^{w}
  +[{\rm Re}(\Pi)]_{vw}E^{w} \label{dualE}
\end{eqnarray}
where the star is the four-dimensional Hodge-star operator and
$\Pi_{vw}$ is the period matrix of the complex curve $\cC_{2}$.
To define this matrix we denote by $(a_{w},b_{w})$ a standard basis of
$H_{1}(\cC_{2})$ consisting of $\a$ and $\b$ cycles
and introduce a set of one-forms $(\a_{w})$ satisfying
$\int_{a_{u}}\a_{w}=\d_{uw}$. Then the period matrix is given by
\begin{equation}
  \Pi_{uw}\equiv\int_{b_{u}}\a_{w}\label{period}.
\end{equation}
For the case of a torus, $g=1$, the period matrix is simply a complex
number which can be identified with the complex structure $\t$ of the
torus. Shortly, we will use the relations~\eqref{dualj} and
\eqref{dualE} to eliminate half of the vector fields as well as
$B_{\m\n}$ in favour of $s$ from our low-energy effective action to
arrive at a description in terms of independent fields.

The remaining bosonic world-volume field we should consider is the
auxiliary scalar field $a$. If we want the normal vector $v$ to be
globally well defined, we cannot allow it to point into the internal
directions of the two-cycle only. This is because generally there need
not exist a nowhere vanishing vector field on a Riemann surface, as
the simple example of a sphere $S^{2}$ already demonstrates.  Hence,
we will take $a$ to be independent of the internal coordinates and
require it to be a function of the external coordinates only, that is,
$a=a(\s^{\m})$. It turns out that this field will drop out of the
five-dimensional effective action after eliminating half of the
degrees of freedom~\footnote{We would like to thank Dmitri Sorokin for
a helpful discussion on this point.}, using Eqs.~\eqref{dualj} and
\eqref{dualE}.

The last ingredient we need to discuss is the non-zero
mode~\eqref{Gsol}. It consists of the purely internal part of the
four-form gauge field strength, but since we now allow the five-brane to
fluctuate it must be slightly generalised. To this end we
define the function
\begin{equation}\label{alpha}
  \a =\a_{1}\th(M_4^1)+\a_{2}\th(M_4^2)+\a_{5}\left[\T(M_{4}^5)+
        \T(\tilde{M}_{4}^5)\right]
\end{equation}
where $d\T (M_4^k)=\d (M_4^k)$ and the $\th$-- and $\d$--function are
defined in analogy with Eqs.~\eqref{Theta} and \eqref{delta}. The
non-zero mode can then be written as
\begin{equation}\label{modNZ}
  G=-\frac{1}{3}\a(x)*\o.
\end{equation}
Note that for the static brane configuration~\eqref{emb} which implies
$\T(M_{4}^5)=\th(y-y_{5})$ and $\T(\tilde{M}_{4}^5)=\th(y+y_{5})$ the
above expression reduces to the background configuration~\eqref{Gsol}
as it should.


\section{The five-dimensional theory}

Based on the above background solutions, we would now like to derive the
five-dimensional effective action and discuss its properties.

\subsection{The action}

Let us start by explaining how the previously identified moduli
fields fit into super-multiplets. In the bulk, we have $D=5$, $N=1$
supergravity with a gravity multiplet $(g_{\a\b},\cA_\a,\Psi_\a^i)$
consisting of the graviton, the gravi-photon $\cA_\a$ with field
strength $\cF_{\a\b}=(d\cA)_{\a\b}$ and the gravitino
$\Psi_\a^i$. Five-dimensional fermions are described by symplectic
Majorana spinors carrying $SU(2)$ R-symmetry indices $i,j,\ldots =
1,2$. The other bulk fields arrange themselves into the universal
hyper-multiplet containing the fields $(V,\s ,\x ,\z^i)$. Here $\s$ is
the dual of the five-dimensional three-form $C_{\a\b\g}$ with field
strength $G_{\a\b\g\d}$. The field strength of the complex scalar $\x$
is denoted by $\cX_\a=d\x$ and $\z^i$ are the fermions. We note that,
from their 11-dimensional origin, the metric and $V$ are
$\mathbb{Z}_2$--even fields, while $C_{\a\b\g}$, $\cA_\a$ and $\x$ are
$\mathbb{Z}_2$--odd.

On the four-dimensional fixed planes $M_4^k$, where $k=1,2$, we have
$N=1$ gauge multiplets, that is gauge fields $A_{k\a}$ with field
strengths $F_{k\a\b}=(dA_k)_{\a\b}$ and the corresponding gauginos. In
general, there will also be gauge matter fields in $N=1$ chiral
multiplets but we will set these fields to zero for simplicity. They
will, however, be included in Ref.~\cite{Inprep}.

On the three-brane world-volume $M_4^5$, the embedding coordinates
$X^\a$ give rise to a single physical degree of freedom $Y=X^{11}$, as
can be seen from the static gauge choice. This field is part of the
$N=1$ chiral multiplet with bosonic content $(Y,s)$. We recall that
the scalar $s$ with field strength $j_\a=(ds-\hat{\cA})_\a$ originates
from the five-brane two-form. We denote the corresponding fermions by
$\th^i$.  In addition, we have $N=1$ gauge multiplets containing
Abelian gauge fields $D_\a^u$ with field strenghts $E_{\a\b}^u$, where
$u,v,w,\ldots =1,\ldots ,g$ and $g$ is the genus of the curve $\cC_2$
wrapped by the five-brane. In general, there will be additional chiral
multiplets parametrizing the moduli space of the five-brane curves
$\cC_2$ but they will not be explicitly taken into account here.

\vspace{0.4cm}

The reduction to five dimensions is not completely straightforward
particularly when dealing with the Chern-Simons and Dirac-term in the
eleven-dimensional action. We have, therefore, performed the reduction
on the level of the equations of motion. This gives rise to the following
effective five-dimensional action
\begin{equation}\label{THEaction}
  S_5=S_{\rm grav}+S_{\rm hyper}+S_{\rm bound}+S_{\rm 3-brane}
\end{equation}
where
\begin{alignat}{1}
  S_{\rm grav}&=-\frac{1}{2\k_{5}^{2}}\int_{M_{5}}\left\{d^{5}x\sqrt{-g}
            \left(\frac{1}{2}R+\frac{3}{2}\cF_{\a\b}\cF^{\a\b}\right)
            +4\cA\w\cF\w\cF\right\}\label{grav}\\
  S_{\rm hyper}&=-\frac{1}{2\k_{5}^{2}}\int_{M_{5}}\bigg\{d^{5}x\sqrt{-g}
            \bigg(\frac{1}{4}V^{-2}\pt_{\a}V\pt^{\a}V+2V^{-1}\cX_{\a}
            \bar{\cX}^{\a}
            +\frac{1}{4!}V^{2}G_{\a\b\g\d}G^{\a\b\g\d}\nn\\
             &\qquad\qquad\qquad\qquad\qquad\qquad +\frac{1}{3}
            V^{-2}\a^{2}\bigg)+2G\w\left(i(\xi\bar{\cX}-\bar{\xi}\cX)
            -2\a\cA\right)
            \bigg\} \label{hyper}
\end{alignat}
\begin{alignat}{1}
  S_{\rm bound}&=-\frac{1}{2\k_{5}^{2}}\left\{2\int_{M_{4}^{1}}d^{4}x
             \sqrt{-g_{4}}
             V^{-1}\a_{1}+2\int_{M_{4}^{2}}d^{4}x\sqrt{-g_{4}}V^{-1}
             \a_{2}\right\}\nn\\
             &\;-\frac{1}{16\pi\a_{\rm GUT}}\left\{\int_{M_{4}^{1}}d^{4}x
             \sqrt{-g_{4}}\, V\,\mbox{tr}(F_{1\mu\n}F_1^{\mu\nu})+
             \int_{M_{4}^{2}}d^{4}x\sqrt{-g_{4}}\, V
             \,\mbox{tr}(F_{2\mu\n}F_2^{\mu\nu})\right\}\label{bound}\\
  S_{\rm 3-brane}&=-\frac{1}{2}T_{3}\Bigg\{\int_{M_4^5\cup\tilde{M}_4^5}
                   \bigg[ d^{4}x\sqrt{-\g}\Big(V^{-1} +2
                   V^{-1}j_{\mu}j^{\mu} +[{\rm Im}(\Pi)]_{uw}E^{u}_{\mu\nu}
                   E^{w\mu\nu}\Big)\nn\\
                   &\qquad\qquad\qquad\qquad\qquad\qquad
                   -4\hat{C}\w ds-2[{\rm Re}(\Pi)]_{uw}E^{u}\w
                   E^{w}\bigg] \Bigg\} \label{3brane}\; .
\end{alignat}
The five-dimensional Newton constant $\k_{5}$, the three-brane tension
$T_{3}$ and the gauge coupling constant $\a_{\rm GUT}$ are given by
\begin{equation}
  \k_{5}^{2}=\frac{\k_{11}^{2}}{v}\, ,\qquad\quad
  T_{3}=\frac{\a_{5}}{\k_{5}^{2}}\, ,\qquad\quad 
  \a_{\rm GUT}=\frac{\l^2}{4\p v}\, .
\end{equation}
In this action all topological terms are written in differential form
whereas all other contributions are given in component form. The hat
denotes the pull-back of a bulk antisymmetric tensor field to the
three-brane world-volume. The induced metric $\g_{\m\n}$ on the
three-brane world-volume is, as usual, defined
by
\begin{equation}
 \g_{\m\n} = \pt_\m X^\a\pt_\n X^\b g_{\a\b}\; .
\end{equation}
The field strength $G$ satisfies the non-trivial Bianchi identity
\begin{equation}
 G = dC - \o_{\rm YM}
\end{equation}
with the Yang-Mills Chern-Simons form $\o_{\rm YM}$ defined by the relation
\begin{equation}
 J_{\rm YM}\equiv d\o_{\rm YM} = \frac{\k_5^2}{4\p\a_{\rm GUT}}\left(
                                 {\rm tr}(F_1\wedge F_1)\wedge\d (y) +
                                 {\rm tr}(F_2\wedge F_2)\wedge\d (y-\p\r )
                                 \right)\, .
\end{equation}
Unlike its 11-dimensional counterpart~\eqref{BI}, this Bianchi has
only contributions from the orbifold planes since, in five dimensions,
the bulk three-brane cannot provide a magnetic source for a four-form
field strength.  We note, that the Bianchi identities for $\cF$ and
$\cX$ also become non-trivial once gauge-matter fields on the orbifold
planes are taken into account~\cite{Lukas:1999tt}. The matrix $\P$
specifying the gauge-kinetic function on the three-brane is the period
matrix defined in Eq.~\eqref{period}. We recall that the
step-function $\a$ in the above action has been defined in
Eq.~\eqref{alpha} and that the charges $\a_k$ satisfy the cohomology
condition
\begin{equation}
 \a_1+\a_2+\a_5 = 0\; .
\end{equation}
Finally, all higher-curvature terms have been dropped from the above action.

\vspace{0.4cm}

Let us now discuss a few elementary properties of the above action. By
construction, this action must represent the bosonic part of a
five-dimensional $N=1$ supergravity theory on the orbifold
$S^1/\mathbb{Z}_2$ coupled to two four-dimensional $N=1$ theories on
the orbifold fixed planes and an additional $N=1$ supersymmetric
three-brane. First we note that once the three-brane is taken away
this action reduces, up to rescalings\footnote{The rescalings are
$C'=\frac{1}{6}2^{1/6}C$, $G'=2^{1/6}G$, $\xi'=2^{1/6}\xi$,
$\cA'=2^{1/6}\cA$, $V'=2^{-2/3}V$ and $g'_{5\a\b}=2^{-2/3}g_{5\a\b}$
where the prime denotes the fields as in Ref.~\cite{Lukas:1999yy}.},
exactly to the action of Ref.~\cite{Lukas:1999yy}, as it should. The
explicit proof that the bulk theory has indeed the correct
supergravity structure can be carried out as in
Ref.~\cite{Lukas:1999yy,Lukas:1999tt} by dualising the three-form
$C_{\a\b\g}$ to a scalar $\s$. This scalar together with its
super-partners $V$ and $\x$ can then be shown to parametrise the
standard universal hyper-multiplet coset $SU(2,1)/SU(2)\times
U(1)$. Moreover, the shift-symmetry of the dilatonic axion $\s$ is
gauged with charge $\a$ and the gravi-photon $\cA$ as the
corresponding gauge boson. This can be directly seen from the term $\a
G\wedge\cA$ in Eq.~\eqref{hyper}. Unlike in the case without
five-branes the gauge charge changes across the bulk as anticipated in
Ref.~\cite{Lukas:1999hk}. From the definition of $\a$,
Eq.~\eqref{alpha}, this charge is proportional to $\a_1$ between the
first fixed plane and the three-brane and proportional to $\a_1+\a_5$
between the three-brane and the second fixed plane. By supersymmetry,
the presence of the bulk potential $\a^2 V^{-2}$ is directly related
to the gauging. Note that, similarly to the gauge charge, this
potential jumps across the three-brane. Further, it is worth pointing
out that, while all tension terms are proportional to $V^{-1}\a_k$,
where $k=1,2,5$, the terms on the fixed planes contain an additional
factor of two relative to the three-brane term. This factor reflects
the nature of the ``boundary branes'' as being located on
$\mathbb{Z}_2$ orbifold fixed planes.

\subsection{Symmetries}

Let us start with the supersymmetry transformations of the fermions.
Their bosonic part can be obtained by either a reduction from 11 dimensions
or, most easily, by generalizing the results of Ref.~\cite{Lukas:1999yy}.
The latter simply amounts to substituting the function $\a$,
Eq.~\eqref{alpha} into the ``massive'' terms in the transformations.
The result is
\begin{eqnarray}
   \d \psi_\a^i &=& D_\a\e^i 
     + \frac{i}{4}
          \left({\g_\a}^{\b\g}-4\d_\a^\b\g^\g\right)\cF_{\b\g}\e^i
     - \frac{1}{\sqrt{2}}V^{-1/2}\left(
        \partial_\a\x\,{(\t_1-i\t_2)^i}_j
        - \partial_\a{\bar\x}\,{(\t_1+i\t_2)^i}_j \right) \e^j
     \nn \\ &&
     - \frac{i}{2\cdot 4!}V{\e_\a}^{\b\g\d\e}\cG_{\b\g\d\e}{(\t_3)^i}_j\e^j
     + \frac{1}{6}\a(x) V^{-1}\g_\a{(\t_3)^i}_j\e^j 
     \label{psi5} \\
 \d\z^i &=& \frac{1}{4!}V\e^{\a\b\g\d\e}\cG_{\a\b\g\d}\g_\e\e^i
     - \frac{i}{\sqrt{2}}V^{-1/2}\g^\a\left(
        \partial_\a\x\,{(\t_1-i\t_2)^i}_j
        + \partial_\a{\bar\x}\,{(\t_1+i\t_2)^i}_j \right) \e^j
     \label{susy5} \nn \\ &&
     + \frac{i}{2}V^{-1}\g_\b\partial^\b V\e^i
     + i\a(x) V^{-1}{(\t_3)^i}_j\e^j \label{z5}
\end{eqnarray}
where $\g_\a$ are the five-dimensional gamma matrices and $\t_i$ are
the Pauli matrices. The variation of the three-brane world-volume spinors
$\th^i$ (assuming the world-volume fields $s$ and $D^u$ vanish) can be
obtained by reducing the variation~\eqref{theta} which results in
\begin{equation}
 \d\th^i = \e^i + {(p_+)^i}_j\k^j \label{th5}
\end{equation}
where the projection operators $p_\pm$ are now given by
\begin{equation}
 p_\pm = \frac{1}{2}\left( 1\pm\frac{i}{4!}\e^{\m_1\ldots\m_4}
         \pt_{m_1}X^{\a_1}\ldots\pt_{m_4}X^{\a_4}\g_{\a_1\ldots\a_4}
         \t_3\right)\; .
\end{equation}

\vspace{0.4cm}

Up to total derivatives the action \eqref{THEaction} is also invariant
under the following gauge variations
\begin{equation}
  \d c=d\l^{(2)},\quad \d \cA=d\l^{(0)},\quad \d s=\hat{\l}^{(0)},\quad
  \d \xi=\mbox{const}.,\quad \d E^{u}=d\l^{(1)u}
\end{equation}
with $\l^{(k)}$ being k-form gauge parameters and $\l^{(0)}$,
$\l^{(2)}$ being $\mathbb{Z}_{2}$--odd. To check this result one must
note that the variation of the gauge term $\sim 4G\w\a(x)\cA$ and the
brane term $\sim 4\hat{C}\w ds$ cancel each other after partial
integration of the former. The above gauge variations of course also
follow from the reduction of the D=11 gauge symmetries
\eqref{gausym1}. Note, however, that there are no remnants of the
PST-symmetries~\eqref{PST} in our action. This is not surprising since
these symmetries have been implicitly gauge-fixed when the self-duality
relations~\eqref{dualj} and \eqref{dualE} were used to eliminate half
of the degrees of freedom on the three-brane.

\subsection{The dual form of the action}

In our five-dimensional action $S_5$, Eq.~\eqref{THEaction}, the
three-brane is coupled to the gauge charge $\a$ defined by
Eq.~\eqref{alpha}. We can promote $\a$ to a zero-form field strength
which satisfies the Bianchi identity
\begin{equation}\label{BIalpha}
  d\alpha=2\a_{1}\d(M_{4}^{1})+2\a_{2}\d(M_{4}^{2})+\a_{5}\left[
    \d(M^{5}_{4})+\d(\tilde{M}^{5}_{4})\right]\; .
\end{equation}
The three-brane then couples magnetically to this zero-form. 
In analogy with massive IIA supergravity~\cite{Bergshoeff:1996ui}, there
should now be a dual formulation of the action $S_5$ where $\a$ is replaced
by a four-form $N$ with five-form field strength $M=dN$ to which the
three-brane couples electrically. If a dual version of 11-dimensional
supergravity involving only a six-form field existed we could have
derived this dual five-dimensional action directly from 11
dimensions. Such a dual version of 11-dimensional supergravity is not
available and, hence, the reduction necessarily leads to the
five-dimensional action $S_5$ written in terms of $\a$. However, there
is no obstruction performing the dualisation in five dimensions. This
can be done by adding to the action $S_5$ (with $\a$ interpreted as a
zero-form field strength) the terms
\begin{equation}
  S_{\a}=\frac{1}{2\k_{5}^{2}}\left\{\int_{M_{5}}N d\a
        -\sum_{k=1}^{2}\int_{M_{4}^{k}}2\a_{k}\hat{N}
        -\a_{5}\left[\int_{M^{5}_{4}}\hat{N}+\int_{\tilde{M}^{5}_{4}}
        \hat{N}\right]\right\}\, .\label{Sa}
\end{equation}
The equation of motion for $N$ now precisely yields the Bianchi
identity~\eqref{BIalpha} for $\a$ by virtue of which the additional
terms~\eqref{Sa} vanish. As it should, this leads us back to the original
action $S_5$ with $\a$ being defined by Eq.~\eqref{alpha}.
On the other hand, the equation of motion for $\a$ computed from
the action $S_5+S_\a$ is given by
\begin{equation}
 \a=-\frac{3}{2}V^{2}*(M-4G\w\cA)\; .
\end{equation}
Using this relation to replace $\a$ in favour of $M$, we arrive at the
dual version of our five-dimensional brane-world action~\eqref{THEaction}.
It is given by
\begin{equation}\label{dualaction}
  S_{\rm 5,dual}=S_{\rm grav}+S_{\rm hyper}+S_{\rm bound}+S_{\rm 3-brane}
\end{equation}
where
\begin{alignat}{1}
  S_{\rm grav}&=-\frac{1}{2\k_{5}^{2}}\int_{M_{5}}\left\{d^{5}x\sqrt{-g}
            \left(\frac{1}{2}R+\frac{3}{2}\cF_{\a\b}\cF^{\a\b}\right)
            +4\cA\w\cF\w\cF\right\}\\
  S_{\rm hyper}&=-\frac{1}{2\k_{5}^{2}}\int_{M_{5}}\bigg\{d^{5}x\sqrt{-g}
            \bigg(\frac{1}{4}V^{-2}\pt_{\a}V\pt^{\a}V+2V^{-1}\cX_{\a}
            \bar{\cX}^{\a}+\frac{1}{4!}V^{2}G_{\a\b\g\d}G^{\a\b\g\d}\\
             &\qquad\qquad +\frac{3}{4\cdot5!}V^{2}M_{\a_{1}\ldots
            \a_{5}}M^{\a_{1}\ldots\a_{5}}\bigg)
            +2G\w\left(i(\bar{\xi}\cX-\xi\bar{\cX})+3V^{2}\cA\w 
            *(M-2G\w\cA)\right)\bigg\} \nn \\
  S_{\rm bound}&=-\frac{1}{2\k_{5}^{2}}\sum_{k=1}^{2}2\a_{k}\int_{M_{4}^k}
           \left\{d^{4}x\sqrt{-g_{4}}V^{-1}+\hat{N}\right\}\\
            &\;-\frac{1}{16\pi\a_{GUT}}\left\{\int_{M_{4}^{1}}d^{4}x
            \sqrt{-g_{4}}V\,{\rm tr}(F_{1\mu\n}F_1^{\mu\nu})+
            \int_{M_{4}^{2}}d^{4}x\sqrt{-g_{4}}V\,{\rm tr}(F_{2\mu\n}
            F_2^{\mu\nu})\right\}\\
  S_{\rm 3-brane}&=-\frac{1}{2}T_{3}\Bigg\{\int_{M_{4}^5\cup\tilde{M}_4^5}
    \bigg[ d^{4}x\sqrt{-\g}\Big(V^{-1} +2V^{-1}j_{\mu}j^{\mu}
    +[{\rm Im}(\Pi)]_{uw}E^{u}_{\mu\nu}E^{w\mu\nu}\Big)\\
    &\qquad\qquad\qquad\qquad\qquad\qquad
     -4\hat{C}\w ds-2[{\rm Re}(\Pi)]_{uw}E^{u}\w
     E^{w}+\hat{N}\bigg] \Bigg\} \nn
\end{alignat}


\section{The vacuum solution}\label{sectsol2}

In this section, we will review the BPS domain-wall solution of
the five-dimensional theory~\eqref{THEaction}. This vacuum state is
associated with an effective $N=1$ four-dimensional theory which we will,
in part, determine explicitly.

\subsection{The supersymmetric domain-wall vacuum state}

For the case without additional bulk three-branes, the supersymmetric
domain-wall solution of five-dimensional heterotic M-theory has
been found in Ref.~\cite{Lukas:1999yy}. We now wish to verify that
this result can be extended to include the effect of the bulk
three-brane thereby providing a solution of our action~\eqref{THEaction}.
We start with the following Ansatz for metric and the dilaton
\begin{equation}
  ds_{5}^{2}=a^{2}(y)\eta_{\mu\nu}dx^{\mu}dx^{\nu}+b^{2}(y)dy^{2}\, ,
  \qquad\quad V=V(y) \label{ansatz}
\end{equation}
with all other bulk fields vanishing. We have to supplement this Ansatz
by a configuration for the three-brane world-volume fields. This
simply corresponds to a static three-brane parallel to the orbifold
planes, that is
\begin{equation}
  X^{\mu}=\s^{\mu},\qquad Y=\mbox{const},
\end{equation}
with all other world-volume gauge fields vanishing. The $E_{8}$ gauge
fields on the boundaries are turned off as well.

With this Ansatz one indeed arrives at an exact solution of the
action~\eqref{THEaction} where $a$, $b$ and $V$ are explicitly given
by
\begin{eqnarray}\label{5dimsol}
  a &=&a_{0}h^{1/2}(y)\nn\\
  b&=&b_{0}h^{2}(y) \\
  V&=&b_{0}h^{3}(y)\nn 
\end{eqnarray}
and the function $h$ has been defined in Eq.~\eqref{bfun}. In
particular, one can check that the new features arising from the
presence of the three-brane are properly taken into
account. Specifically, the three-brane world-volume equations of motion
are satisfied and the three-brane sources in the Einstein equation are
properly matched. This solution represents a triple domain-wall
matching not only the sources on the orbifold fixed planes but, in
addition, it matches the bulk three-brane source and its mirror
source. This can be explicitly seen from the function $h$ which
satisfies
\begin{equation}
  \pt^{2}_{y}h=-\frac{2}{3}\pt_{y}\a(y)=-\frac{4}{3}\left[\a_{1}
               \d(y)+\a_{2}\d(y-\p\r)+\frac{1}{2}\a_{5}\left(\d(y-Y)
               +\d(y+Y)\right)\right]\; .
\end{equation}
The $\d$--functions indicate the positions of the various orbifold planes/
three-branes at $y=0,\p\r ,Y,-Y$.

This solution is also a BPS state of the theory since it preserves
four of the eight supersymmetries. This can be verified by
using Eqs.~\eqref{psi5} and \eqref{z5} for the bulk fermions and
Eq.~\eqref{th5} with $\k^i=-\e^i$ for the three-brane world-volume
fermions. The Killing spinor is explicitly given by
\begin{equation}
  \e^{i}=H^{1/4}\e_{0}^{i},\quad \g_{11}\e_{0}^{i}=
         (\t_{3})^{i}_{\,j}\e^{j}_{0},
\end{equation}
where $\e_{0}^{i}$ is a constant spinor.

\subsection{The four-dimensional effective theory}

The above domain-wall vacuum state is associated with an $N=1$
effective four-dimensional theory describing fluctuations
around this state. We would now like to compute some aspects
of this four-dimensional theory.

The bosonic moduli fields from the bulk are the four-dimensional
metric $g_{\m\n}$, the Calabi-Yau volume $V$ and the orbifold size
$R=\sqrt{g_{55}}$ (both averaged over the orbifold), the axion
$\c=\cA_5$ and the two-form $B_{\m\n}=C_{11\m\n}$. The latter can, in
four dimensions, be dualised to a scalar $\s$. From the three-brane
world-volume, we have the scalar $z=Y/\p\r\in [0,1]$ specifying the
position of the three brane and the axion $\n = s/\p\r$ together with
the $g$ Abelian gauge fields $D^u$ with field strengths $E^u$, where
$u,v,w,\ldots = 1,\ldots ,g$. Recall that $g$ is the genus of the
curve $\cC_2$ within the Calabi-Yau space which is wrapped by the
five-brane. Finally, we should consider the two gauge fields $A_k$
with field strengths $F_k$, where $k=1,2$, which originate from the
orbifold fixed planes. Obviously, all gauge fields fall into $N=1$
gauge multiplets.

The six scalar fields, on the other hand, fit into three chiral
multiplets, namely the dilaton $S$, the $T$-modulus and the
five-brane modulus $Z$. The bosonic parts of these fields turn out
to be related to the component fields by
\bea
 S &=& V+q_5Rz^2+i(\s +2q_5\c z^2) \nn \\
 T &=& R + 2i\c \label{comp}\\
 Z &=& Rz+2i (-\n +z\c )\nn
\eea
where
\begin{equation}
 q_k = \p\r\a_k
\end{equation}
for $k=1,2,5$. Recall that $\a_5$ is the five-brane charge defined
in Eq.~\eqref{alphak}.
The moduli K\"ahler potential for the superfields $S$, $T$ and $Z$
can now be found by a reduction of the five-dimensional
action~\eqref{THEaction} on the domain-wall vacuum state. This leads to
\begin{equation}
 K_{\rm moduli} = -\ln\left[ S+\bar{S}-q_5\frac{(Z+\bar{Z})^2}{T+\bar{T}}
                  \right] - 3\ln\left[ T+\bar{T}\right]\; ,\label{K}
\end{equation}
confirming the earlier result~\cite{Derendinger:2001gy}
which was obtained by somewhat different
methods. We note that the component form~\eqref{comp} of the
superfields allows a direct interpretation of this K\"ahler potential
and the resulting moduli field dynamics in five-dimensional
terms. From the non-trivial structure of the domain-wall, it is
possible to compute loop-corrections of order $\e_S$ to the kinetic
terms of the moduli. However, we did not succeed in finding the
complex structure and the associated corrected K\"ahler potential when
those corrections were included. It is conceivable that this
computation is beyond the range of validity of the five-dimensional
theory. This is supported by the observation that the $Z$--dependent
part in the above K\"ahler potential~\eqref{K} is already suppressed
by $\e_S\sim R/V$ relative to the $S$--dependent part. This suggests
that corrections are already of order $\e_S^2$ and, therefore, beyond
the linear level up to which the five-dimensional theory can
generally be trusted.

The gauge kinetic functions for the gauge fields $A_1$ and $A_2$ from
the orbifold fixed planes turn out to be
\bea
 f_1 &=& S-q_2T-2q_5Z \\
 f_2 &=& S+q_2T\, ,
\eea
again in agreement with
Ref.~\cite{Lukas:1999hk,Derendinger:2001gy}. We note that the
threshold terms proportional to $T$ and $Z$ arise as a direct
consequence of the domain-wall structure. In fact, the correction is
entirely due to the non-trivial orbifold dependence of the
dilaton in Eq.~\eqref{ansatz} since conformal invariance of four-dimensional
Yang-Mills theory implies that the warping of the five-dimensional
metric drops out. As a consequence, no such threshold correction
arises for the three-brane gauge fields, since their kinetic term in
Eq.~\eqref{3brane} does not depend on the dilaton. After an
appropriate rescaling of the fields $D^u$ their gauge-kinetic function
is simply proportional to the period matrix~\eqref{period} of the holomorphic
curve $\cC_2$, that is 
\begin{equation}
 f_{uv}=i\P_{uv}\; .
\end{equation}


\vspace{1cm}

\noindent
{\Large\bf Acknowledgements}\\
A.~L.~is supported by a PPARC Advanced Fellowship. M.~B.~is supported by a
Graduiertenkolleg Fellowship of the DFG.


\end{document}